\documentstyle[osa,manuscript,psfig]{revtex}

\begin{document}

\titlepage
\title{Spin alignment of vector meson in $e^+e^-$ annihilation
 at $Z^0$ pole}
\author{Xu Qing-hua, Liu Chun-xiu and Liang Zuo-tang}
\address{Department of Physics, Shandong University,
Jinan, Shandong 250100, China}
\maketitle

\begin{abstract}
We calculate the spin density matrix of the vector meson produced in 
$e^+e^-$ annihilation at $Z^0$ pole. We show that the data imply 
a significant polarization for the antiquark  
which is created in the fragmentation
process of the polarized initial quark and combines with the fragmenting 
quark to form the vector meson. The direction of polarization
is opposite to that of the fragmenting quark and the magnitude
is of the order of 0.5. A qualitative explanation of this result
based on the LUND string fragmentation model is given.
\end{abstract}

\newpage

Spin effects in high energy fragmentation processes have attracted
much attention [1-11] 
recently since they provide not only useful 
information on the spin dependence of hadronic interaction
but also a promising tool to study the spin structure of nucleon.
One of the important issue in this connection is the 
spin transfer of the fragmenting quark to the produced hadron
which contains this quark. 
There are two related questions here, one is whether the
fragmenting quark retains its polarization, 
the other is how one should connect the spin of the 
fragmenting quark to that of the hadron which contains the quark.
  
A series of papers [1-3, 6-11],
both experimental and theoretical,
have been published on this topic. 
These papers all concentrate on the hyperon polarization
in lepton induced reactions since hyperon polarization
can easily be obtained in experiments by measuring 
the angular distributions of its decay products.
Although the available data are still far from accurate
and enormous enough to make a definite conclusion,
the comparison of the data and theoretical results
seems to suggest that\cite{f15,f5} the polarization of the fragmenting 
quark is retained in the fragmentation and that 
the simple SU(6) wave function can be used 
in connecting the spin of the fragmenting quark
to that of the produced hadron which contains the quark.

It is also interesting to note that information
on polarization of vector meson can also be obtained from
angular distributions of their decay products.
Such measurements have also been carried out
by OPAL, DELPHI and ALEPH collaboration at LEP
for different vector mesons in 
$e^+e^-$ annihilation at $Z^0$ pole [12-16].
The results show clearly that the produced vector meson
favor the helicity zero state 
which implies a nonzero polarizaton of 
the vector meson in the direction
perpendicular to the moving direction.
There are also measurements on the off-diagonal elements 
of the spin density matrix.
It is then natural to ask whether the data can also 
be understood from the same starting points as 
those for hyperon polarization and what they imply
for the polarization of the antiquark produced 
in the fragmentation process of a polarized quark.
                    
In this note, we study these questions by calculating 
the spin density matrix for the produced vector mesons
by adding the spin of the fragmenting quark (or antiquark) with that 
of the antiquark (or quark) created in the fragmentation together.
With the aid of an event generator JETSET, we calculate
the $z$
dependence of the polarization by taking all
different contributions into account. 
Here $z\equiv2p_V/\sqrt{s}$, where $p_V$ is the momentum of 
the vector meson, $\sqrt{s}$ is the total $e^+e^-$ center of mass energy.
We now first outline
the calculation method in the following.

Similar to that for hyperon polarization \cite{f5}, to calculate the
polarization of vector mesons in the inclusive process 
$e^+e^-\rightarrow Z^0 \rightarrow q^0_f\bar{q}^0_f
\rightarrow$ V+X, we need to divide the produced 
vector mesons into the following two groups and consider them separately.
(Here the subscript $f$ denotes the flavor of the quark.)
They are
(a) those which contain the initial quark $q^0_f$'s 
or the initial antiquark $\bar{q}^0_f$'s;
(b) those which don't contain the initial quark or antiquark.
The spin density matrix $\rho^V(z,k_\perp)$ for the vector meson V
should be given by:   
\begin{equation}
\rho^V(z,k_\perp)=\sum_f 
\frac {\langle n(z,k_\perp|a,f)\rangle}{\langle n(z,k_\perp)\rangle}
\rho^V(z,k_\perp|a,f)+
\frac {\langle n(z,k_\perp|b)\rangle}{\langle n(z,k_\perp)\rangle}
\rho^V(z,k_\perp|b)
,
\end{equation}
where $\langle n(z,k_\perp|a,f)\rangle$ and $\rho^V(z,k_\perp|a,f)$ 
are the average number and spin density matrix of vector mesons from (a);
$\langle n(z,k_\perp|b)\rangle$ and $\rho^V(z,k_\perp|b)$ are those from (b).
$\langle n(z,k_\perp)\rangle$ =
$\sum_f \langle n(z,k_\perp|a,f)\rangle$ +
$\langle n(z,k_\perp|b)\rangle$ is the 
total number of vector mesons
and $k_\perp$ is the transverse momentum 
of the vector meson with respect to the moving direction 
of initial quark or antiquark. 
Here, in contrast to the case for $\Lambda$ hyperon production,
contributions from the decay of heavier hadrons are very small.
We just treat them in the same way as those from (b).

Similar to the case for hyperon production,
there are many different possibilities
to produce the vector mesons in group (b) and the polarization in
each possibility can be different from that in the other.
Hence, it is very likely that these vector mesons as a whole 
are unpolarized. We will, just as we did in Ref. \cite{f5} for hyperons,
take them as unpolarized.
This means that we simply take 
$\rho^V(z,k_\perp|b)$ as a unit matrix.
The spin density matrix $\rho^V(z,k_\perp|a,f)$ of the vector meson V
which contains the fragmenting quark $q^0_f$ (or antiquark ${\bar q}^0_f$) 
and an antiquark ${\bar q}$ (or a quark $q$) which is created in the 
fragmentation process can be calculated 
from the direct product of the spin density matrix 
$\rho^{q^0_f}$ for $q^0_f$ (or $\rho^{{\bar q}^0_f}$ for ${\bar q}^0_f$) 
and that $\rho^{\bar{q}}$
for the antiquark $\bar{q}$ (or $\rho^q$ for q).
Now, we are going to formulate the calculation of $\rho^V(z,k_\perp|a,f)$.
For explicity, we will take the case $V=(q^0_f\bar q)$ as an example. 

According to the Standard Model for electro-weak interaction,
the initial quark or antiquark produced in
$e^+e^-$ annihilation at $Z^0$ pole is 
longitudinally polarized and 
the average value of polarization is $P_f=-$0.94 for
$f=d, s$ and $b$; $P_f=-$0.67 for $f=u, c$ \cite{f18,f5}.  
Hence the normalized spin density matrix $\rho^{q^0_f}$ is given by
\begin{equation}
\rho^{q^0_f}= \left (
\begin{array}{cc}
c_{1f}   &   0     \\
0        &   c_{2f}\\ 
\end{array} \right)
,
\end{equation}
where $c_{1f}=(1+P_f)/2$, $c_{2f}=(1-P_f)/2$.
Here we use the helicity frame of $q^0_f$,
which we denote by $oxyz$,
where $z$-axis is taken as the moving direction
of $q^0_f$ in the overall
center-of-mass frame and $x$-axis as the direction of the 
transverse momentum of the antiquark $\bar {q}$
with respect to the $z$-axis.
The $\rho^{\bar q}$ is taken as the most general 
form in the frame $oxyz$, i.e.,
\begin{equation}
\rho^{\bar q} 
=\frac{1}{2} \left (
\begin{array}{cc}
1+P_z    & P_x-iP_y \\
P_x+iP_y & 1-P_z    \\
\end{array} \right)
,
\end{equation}
where $\vec{P}(P_x,P_y,P_z$) is the polarization vector of  
$\bar{q}$.
The spin density matrix $\rho^{q^0_f{\bar q}}$ of 
the $q^0_f{\bar q}$ system is obtained from Eq. (2) and Eq. (3), i.e,
\begin{equation}
\rho^{q^0_f{\bar q}}=\rho^{q^0_f} \otimes \rho^{\bar q}
= \left (
\begin{array}{cc}
c_{1f}\rho^{\bar q} &      0       \\
0                   & c_{2f}\rho^{\bar q} \\ 
\end{array}  \right)
.
\end{equation}

We note that the $\rho^{q^0_f{\bar q}}$
obtained in this way is in the basis of 
$|s^{q^0_f}, s^{q^0_f}_z; s^{\bar q},s^{\bar q}_z\rangle$,
where $s^{q^0_f}$ and $s^{\bar q}$ are the spins of 
$q^0_f$ and $\bar {q}$, and $s^{q^0_f}_z$ and $s^{\bar q}_z$ 
are their $z$ components.
To obtain the spin density matrix for the meson,
we need to transform it to the coupled basis
$|s, s_z\rangle$, where $\vec {s}={\vec {s}^q}+{\vec {s}^{\bar q}}$. 
The bases of these two representations are related to each other
by a unitary matrix U,
\begin{equation}
U= \left (
\begin{array}{cccc}
1   & 0              & 0               & 0  \\
0   &\frac{1}{\sqrt2}&\frac{1}{\sqrt2} & 0  \\
0   & 0              & 0               & 1  \\
0   &\frac{1}{\sqrt2}&-\frac{1}{\sqrt2}& 0
\end{array} \right )
.
\end{equation}
The spin density matrix $\rho^{m}$ in the coupled basis 
is obtained from $\rho^{m}=U\rho^{q^0_f{\bar q}}U^{-1}$, i.e.,
\begin{equation}
\small{
\rho^{m}=
\frac{1}{2} \left (
\begin{array}{cccc}
c_{1f}(1+P_z)                   &
 \frac{c_{1f}}{\sqrt2}(P_x-iP_y)                &
 0                               & 
 \frac{c_{1f}}{\sqrt2}(P_x-iP_y)                \\
\frac{c_{1f}}{\sqrt2}(P_x+iP_y) &
 \frac{1}{2}(1-P_fP_z)      &
 \frac{c_{2f}}{\sqrt2}(P_x-iP_y) &
 \frac{1}{2}(P_f-P_z)\\
0                               &
 \frac{c_{2f}}{\sqrt2}(P_x+iP_y)                &
 c_{2f}(1-P_z)                   &
 -\frac{c_{2f}}{\sqrt2}(P_x+iP_y)               \\
\frac{c_{1f}}{\sqrt2}(P_x+iP_y) &
 \frac{1}{2}(P_f-P_z)        &
 -\frac{c_{2f}}{\sqrt2}(P_x-iP_y)&
 \frac{1}{2}(1-P_fP_z)
\end{array}  \right)
.
}
\label{eq6}
\end{equation}
Hence the spin density matrix $\rho'^V(a,f)$ for the vector meson
V in the basis of $|s^V, s^V_z\rangle$ can be read out 
from Eq. (\ref{eq6}) as follows:
\begin{equation}  
\rho'^V(a,f)=\frac{2}{3+P_fP_z}
\small {
\left (
\begin{array}{ccc}
c_{1f}(1+P_z)                   &
 \frac{c_{1f}}{\sqrt2}(P_x-iP_y)                &
 0                               \\
\frac{c_{1f}}{\sqrt2}(P_x+iP_y) &
 \frac{{1}}{2}(1-P_fP_z)&
 \frac{c_{2f}}{\sqrt2}(P_x-iP_y) \\
0                               &
 \frac{c_{2f}}{\sqrt2}(P_x+iP_y)                &
 c_{2f}(1-P_z)                   \\
\end{array}  \right)
.
}
\end{equation}
In addition, we obtain also the ratio $P/V$ of 
pseudoscalar meson to vector meson from Eq. (\ref{eq6}) as
\begin{equation}
P/V
=\frac {1-P_fP_z}{3+P_fP_z}
.
\label{eq7}
\end{equation} 
We can see that $P/V=1/3$ at $P_z=0$ and
$P/V>1/3$ at $P_fP_z<0$.
    
To compare $\rho'^V(a,f)$ with the data [12-16], we need to 
transform it to the helicity basis of the vector meson,
i.e., the helicity beam frame which we denote by $OXYZ$.
In this frame, $Z$-axis is taken as the moving direction 
of the vector meson and $Y$-axis is 
taken as the vector products of 
$Z$ and the $e^-$ beam direction (see Fig. 1).  
This frame transformation can be carried out by
two successive rotations:
first a rotation of angle $\beta$ around $y$ axis 
which transforms $oxyz$ to $ox'yZ$, 
then a rotation of angle $\gamma$ around the  
$Z$-axis which transforms $ox'yZ$ to $OXYZ$, i.e.,
\begin{equation}
\rho^V(z,k_\perp|a,f)=D^{\dagger}(\beta,\gamma)\rho'^V(a,f)
D(\beta,\gamma)
,
\label{eq8}
\end{equation}
where $D(\beta,\gamma)$ is the rotation matrix
\begin{equation}
D(\beta,\gamma)= \left (
\begin{array} {ccc}
e^{-i\gamma}\cos^2{\frac{\beta}{2}}      &-\frac{1}{\sqrt{2}}\sin{\beta} &
 e^{i\gamma}\sin^2{\frac{\beta}{2}}       \\
\frac{1}{\sqrt{2}}e^{-i\gamma}\sin{\beta}& \cos{\beta}  &
 -\frac{1}{\sqrt{2}}e^{i\gamma}\sin{\beta} \\
e^{-i\gamma}\sin^2{\frac{\beta}{2}}      &\frac{1}{\sqrt{2}}\sin{\beta}  &
 e^{i\gamma}\cos^2{\frac{\beta}{2}}
\end{array} \right)
.
\end{equation}
Here $\beta$ is the angle between the  momentum of the
initial quark $q_f^0$ and that of the vector meson 
and $\gamma$ is the angle between the $y$-axis and $Y$-axis (see Fig. 1).
Clearly, $\beta$, $\gamma$, $z$ and $k_{\perp}$
are related to each other by
$\sin{\beta}=2k_{\perp}/(z\sqrt{s})$ and 
\begin{equation}
\cos{\gamma}=\frac{k_z\cos{\phi}\sin{\theta}
-k_{\perp}\cos{\theta} }
{\sqrt{ k^2_{\perp}\sin^2{\phi}+
(k_z\sin{\theta}-
k_{\perp}\cos{\phi}\cos{\theta})^2 
}}
,
\end{equation}
where $k_z=\sqrt{z^2s/4-k^2_{\perp}}$ and 
$\sin{\gamma}<0$ for $\phi< \pi$ and $\sin{\gamma}>0$ for $\phi> \pi$; 
$\theta$ is the angle between the moving direction of 
the initial quark and the $e^-$ beam direction in
the overall center-of-mass frame and $\phi$ is the azimuthal
angle of $k_{\perp}$ with respect to the plane of the initial
quark and $e^-$ beam.

After some straightforward algebra, we obtain   
the following expressions for the matrix elements which have 
been measured experimentally.
\begin{equation}
\rho_{00}^V(z,k_\perp|a,f)=
\frac{1}{3+P_fP_z}
[1-P_f(P_z\cos{2\beta}+
P_x\sin{2\beta})],
\label{eq00}
\end{equation}
\begin{equation} 
Re \rho_{1,-1}^V(z,k_\perp|a,f)=
\frac{P_f\sin{\beta}}{3+P_fP_z}
[\cos{2\gamma}(P_z\sin{\beta}-
P_x{\cos{\beta}})
-P_y\sin{2\gamma}]
,
\end{equation}
\begin{equation}
Im \rho_{1,-1}^V(z,k_\perp|a,f)=
\frac{P_f\sin{\beta}}{3+P_fP_z}
[\sin{2\gamma}
(P_z\sin{\beta}-
P_x\cos{\beta})
+P_y\cos{2\gamma}]
,
\end{equation}
\begin{eqnarray*}
Re [\rho_{1,0}^V(z,k_\perp|a,f)-\rho_{0,-1}^V(z,k_\perp|a,f)]
\end{eqnarray*}
\begin{equation}
\phantom{1234567}
=\frac{\sqrt{2}P_f}{3+P_fP_z}
[\cos{\gamma}(P_x\cos{2\beta}-P_z\sin{2\beta})
+P_y\cos{\beta}\sin{\gamma}]
,
\end{equation}
\begin{eqnarray*}
Im [\rho_{1,0}^V(z,k_\perp|a,f)-\rho_{0,-1}^V(z,k_\perp|a,f)]
\end{eqnarray*}
\begin{equation}
\phantom{1234567}
=\frac{\sqrt{2}P_f}{3+P_fP_z}
[\sin{\gamma}(P_x\cos{2\beta}-P_z\sin{2\beta})
-P_y\cos{\beta}\cos{\gamma}]
.
\end{equation} 

From these results, we see already the following 
qualitative conclusions:

(1) From Eqs. (8) and (\ref{eq00}), we see that not only 
$\rho^V_{00}(z,k_{\perp}|a,f)$ but also $P/V$ depends on
the polarization of the fragmenting quark $q^0_f$ and that of
the antiquark $\bar{q}$ created in the fragmentation process.
There is a simple relation between them, i.e.,
\begin{equation}
\rho_{00}^V(z,k_\perp|a,f)=
(P/V)+
\frac{2P_f\sin{\beta}}{3+P_fP_z}
(P_z\sin{\beta}-P_x\cos{\beta})
.
\end{equation}
We see that the relation is in general dependent
on the momentum of the produced vector meson.
If we take as usual $k_{\perp}$=
0.35 $GeV$, we obtain $\sin{\beta}= 2k_{\perp}/(z\sqrt{s})\approx 
0.0077/z$ for $\sqrt{s}=91 GeV^{-2}$, 
which is much less than 1 for large $z$.  
Hence we have approximately 
\begin{equation}
\rho_{00}^V(z,k_\perp|a,f)\approx P/V
.
\end{equation}
For $P_z=P_x=0$, we have $\rho_{00}^V(z,k_\perp|a,f)=P/V=1/3$.
This is the result that is expected \cite{spin13} in the unpolarized case. 

(2) From Eqs. (13-16), we see that the non-diagonal
elements $\rho^V_{1,-1}$ and $(\rho_{1,0}^V-\rho_{0,-1}^V)$
depend not only on $\beta$
but also on $\gamma$  which is a function of $\phi$
and other variables. Since for a given $k_\perp$ and 
a given $z$, $\phi$ is distributed uniformly, we should
average over $\phi$ for these quantities.
For $k_\perp \ll (z\sqrt{s}/2)$, we expand $\cos{\gamma}$
and $\sin{\gamma}$ in terms of $k_{\perp}/k_z$ and
keep only terms up to $k_{\perp}/k_z$, 
then average over $\phi$, we obtain
$\langle \cos{\gamma}\rangle \approx
-{\tan{\beta}\cot{\theta}}/{2}$; 
$\langle \sin{\gamma}\rangle\approx 0$;
$\langle \cos{2\gamma}\rangle \approx
-\tan^2{\beta}$ and 
$\langle \sin{2\gamma}\rangle\approx 0$.
We insert them into Eqs. (13-16) and obtain
\begin{equation}
Re \rho_{1,-1}^V(z,k_\perp|a,f) \approx
-\frac{P_f\sin{\beta}}{3+P_fP_z}
\tan^2{\beta}(P_z\sin{\beta}-
P_x{\cos{\beta}})
,
\end{equation}
\begin{equation}
Im \rho_{1,-1}^V(z,k_\perp|a,f)\approx -
\frac{P_f\sin{\beta}}{3+P_fP_z}
P_y\tan^2{\beta}
,
\end{equation}
\begin{equation}
Re [\rho_{1,0}^V(z,k_\perp|a,f)-\rho_{0,-1}^V(z,k_\perp|a,f)]
\approx -\frac{\sqrt{2}P_f\tan{\beta}\cot{\theta}}{2(3+P_fP_z)}
(P_x\cos{2\beta}-P_z\sin{2\beta})
,
\end{equation}
\begin{equation}
Im [\rho_{1,0}^V(z,k_\perp|a,f)-\rho_{0,-1}^V(z,k_\perp|a,f)]
\approx \frac{\sqrt{2}P_fP_y\sin{\beta}\cot{\theta}}{2(3+P_fP_z)}
.
\end{equation}
We see that they contain at least one  
$\sin{\beta}$ which is of the order of $0.01$ at large $z$.
This shows that these non-diagonal elements obtained in this way are
all very small. More precisely, both $Re\rho_{1,-1}$ and $Im\rho_{1,-1}$
should be smaller than $10^{-6}$ and $Re [\rho_{1,0}-\rho_{0,-1}]$ and 
$Im[\rho_{1,0}-\rho_{0,-1}]$ should be small than $10^{-2}$.
Significant nonzero results of these elements may imply significant
interferences \cite{Anselmino} between the fragmentation of 
$q_f^0$ and that of ${\bar{q}}_f^0$, which are not taken into account here.
Experimental results \cite{f8,f9,f2} from OPAL and DELPHI for $Im \rho_{1,-1}$,
$Re [\rho_{1,0}-\rho_{0,-1}]$ and $Im [\rho_{1,0}-\rho_{0,-1}]$ 
seem to show no deviation from the above-mentioned expectations,
there is however signature of deviation for $Re\rho_{1,-1}$,
but the statistics are still too low to make any definite conclusions.

(3) From Eq. (\ref{eq00}), we see that if $P_z=0$, i.e., the
$\bar {q}$ is unpolarized in the moving direction of 
$q_f^0$, we obtain that
\begin{equation}
\rho_{00}^V(z,k_\perp|a,f)|_{P_z=0}=
(1-P_fP_x\sin{2\beta})/3
.
\end{equation}
As we discussed above, $\sin{2\beta}\ll 1$ for large $z$, 
so that 
$\rho_{00}^V(z,k_\perp|a,f)|_{P_z=0}\approx 1/3$.
This shows clearly that $P_z\neq 0$ is a necessary condition for
$\rho^V_{00}\neq 1/3$ at large $z$.
Furthermore, neglecting terms proportional to $\sin{\beta}$ in Eq. (12), 
we obtain, 
\begin{equation}
\label{r0026}
\rho_{00}^V(z,k_\perp|a,f)\approx
{(1-P_fP_z)}/{(3+P_fP_z)}
.
\end{equation}   
This shows that $\rho_{00}^V(z,k_\perp|a,f)> 1/3$
if $P_f$ and $P_z$ have opposite sign.
Both OPAL and DELPHI data [13-16] explicity show that 
$\rho^V_{00}> 1/3$ for all the vector mesons except for $\omega$. 
This implies that $P_z\neq 0$
and has the opposite sign as $P_f$ in these cases.

After we obtain the results for $\rho^V(z|a,f)$ and $\rho^V(z|b)$, 
we can calculate $\rho^V(z)$ if we know the average numbers
$\langle n(z|a,f)\rangle$ and $\langle n(z|b)\rangle$. 
These average numbers are
determined by the hadronization mechnism and should be 
independent of the polarization of the initial quarks.
Hence, we can calculate them using a hadronization model 
which gives a good description of the unpolarized data
for mutiparticle production in high energy reactions.
Presently, such calculation can only be carried out 
using a Monte-Carlo event generator. We use the LUND string 
fragmentation model \cite{lund} implemented by JETSET \cite{jet} 
to do this calculation.
 
Using the event generator JETSET, we calculated
$\langle n(z|a,f)\rangle$ and $\langle n(z|b)\rangle$.
We use Eq. (\ref{r0026}) to calculate $\rho_{00}^V(z,k_\perp|a,f)$
in which the only free parameter is $P_z$.
We first calculate $\rho_{00}$ for $K^{*0}$ as a 
function of $z$ since the corresponding data is available \cite{f9}.
We found out that, to fit the data, the value of $P_z$
has to be quite large. In Fig. 2 we show the results obtained  
by taking $P_z=0.48$.
We see that the data can reasonably well be fitted.

Similiar calculations can certainly be made for other vector
mesons such as $\rho$, $\phi$, $D^{*\pm}$, $B^{*}$ $etc$.
But, for these mesons, only the average values of $\rho_{00}$
in certain $z$ ranges are available. It can easily be 
obtained from Eq. (1),
such average values are given by,
\begin{equation}
\langle\rho^V_{00}\rangle=\sum_f
\frac {\langle n^V(a,f)\rangle}{\langle n^V\rangle}
\rho^V_{00}(a,f)+
\frac {\langle n^V(b)\rangle}{\langle n^V\rangle}
\rho^V_{00}(b)
,
\end{equation}
where $\rho^V_{00}(a,f)$ is given by Eq. (\ref{r0026}) and 
$\rho^V_{00}(b)=1/3$; $\langle n^V(a,f)\rangle$ and $\langle n^V(b)\rangle$
are the number of vector mesons from (a) and (b) in the corresponding
$z$ range respectively and $\langle n^V\rangle$ the total number of vector
meson in that $z$ range.
Using JETSET, we calculate $\langle n^V(a,f)\rangle$ and
$\langle n^V(b)\rangle$.
After that we can determine the $P_z$ which we need to fit
the data on $\langle\rho^V_{00}\rangle$.
We found out that, for most of the vector mesons, 
the resulting $P_z$ can be written as 
$P_z=-\alpha{} P_f$ with $\alpha{}$ is common for most
of the $q_f^0$'s. In Table 1, we show the obtained results by 
choosing $\alpha=0.51$ as determined above from the data
for $K^{*0}$. 
We see that the data can be fitted reasonably well 
except those for $\omega$ and $B^*$. 
The data of $\langle \rho^{\omega}_{00}\rangle$ 
can only be fitted by taking $\alpha$ as negative.
The deviation in the case of $B^*$ may be attributed to 
the helicity flip of $b$-quark caused by gluon radiation, 
which is negligible for light quarks. 

The results except that for $\omega$ can be understood 
qualitatively in the string fragmentation
model. Here, it was shown that \cite{lund} the probability to creat a
meson of mass $m$ is proportional to $e^{{-bm^2}/{x}}$,
where $x$ is the fractional energy of meson from
the fragmenting quark 
and $b$ is a positive parameter in LUND model.
For a polarized quark $q_f^0$ with polarization $P_f$,
if the spin of $\bar{q}$ is in the opposite direction
as that of $q_f^0$, the resulting mesons can be a vector meson
or a pseudoscalar meson with equal probabilities.
The average mass is ${m_1}=(m_V+m_P)/2$,
where $m_V$ and $m_P$ are the masses for the vector meson
and pseudoscalar meson respectively. If the spin of $\bar{q}$ is in
the same direction as that of $q_f^0$, only vector meson can be
created and its mass is $m_2=m_V$.
It is clear that $m^2_1<m^2_2$, thus the corresponding 
probability is large for the former case than that for the latter.
This leads to a longitudinal polarization of $\bar{q}$ and
the polarization is proportional to $P_f$, i.e,
$P_z=-{\alpha}^f_{str}P_f$,
\begin{equation}
{\alpha}^f_{str}=\frac{2}{e^{-b(m^2_2-m^2_1)/x}+1}-1
\end{equation} 
We can see that the sign of $P_z$ is indeed opposite to that of $P_f$. 
In case of $K^{*0}$ which 
contains an initial $d$ or $\bar{s}$ 
we obtain ${\alpha}^f_{str}=0.3$ for $x=0.3$ by taking 
$b=0.58GeV^{-2}$ as was used in JETSET \cite{jet}. 

In summary, we calculated the spin density matrix 
of the vector meson produced in $e^+e^-$ annihilation
at $Z^0$ pole from a direct sum of the 
spin of the polarized fragmenting quark and that of 
the antiquark created in the fragmentation process.
The result for $\rho_{00}$ implies 
a significant polarization for the antiquark which is created
in the fragmentation process and combines with the fragmenting
quark to form the vector meson in the opposite direction as that of the
fragmenting quark. 
The polarization can approximately be written as $P_z=-\alpha P_f$ and 
$\alpha\approx 0.5$ for most of the mesons.

We thank Li Shi-yuan, Xie Qu-bing and other members of the 
theoretical particle physics in Shandong for 
helpful discussions. This work was supported in part by
the National Science Foundation of China (NSFC) and 
the Education Ministry of China.

\newpage

\begin{table}[h]
\caption{Spin density matrix element $\rho_{00}$ 
for different vector mesons obtained from Eq.(25) 
using $P_z=-0.51P_f$.
The data are taken from Refs.[\ref{f4}-\ref{f3}]. }
\begin{center}
\begin{tabular}{cccc} \hline
 meson&$\rho_{00}$  & data           &$z$ range       \\ \hline
$\rho^{\pm}$&0.398 & $0.373\pm0.052$(OPAL)&$0.3<z<0.6$    \\ \hline
$\rho^0$    &0.428 & $0.43\pm0.05$(DELPHI)&$z>0.4$        \\ \hline
$\omega$    &0.405 & $0.142\pm0.114$(OPAL)&$0.3<z<0.6$    \\ \hline
$K^{*0}$    &0.504 & $0.46\pm0.08$(DELPHI)&$z>0.4$        \\ \hline
$\phi  $    &0.557 & 
   \begin{minipage}{4.6cm}
    \begin{center}
    $0.54\pm0.06\pm0.05$(OPAL)\\
    $0.55\pm0.10$(DELPHI)
    \end{center}
   \end{minipage}  &$z>0.7$ \\ \hline
$D^{*\pm}$ &0.415& $0.40\pm0.02\pm0.01$(OPAL) &$z>0.5$  \\ \hline
$B^*$      &0.567& 
   \begin{minipage}{4.6cm}
    \begin{center}
    $0.32\pm0.04\pm0.03$(DELPHI)\\
    $0.33\pm0.06\pm0.05$(ALEPH)\\
    $0.36\pm0.06\pm0.07$(OPAL)
    \end{center}
   \end{minipage}
&$0<z<1$ \\ \hline
\end{tabular}
\end{center}
\end{table}

\begin{figure}
\psfig{file=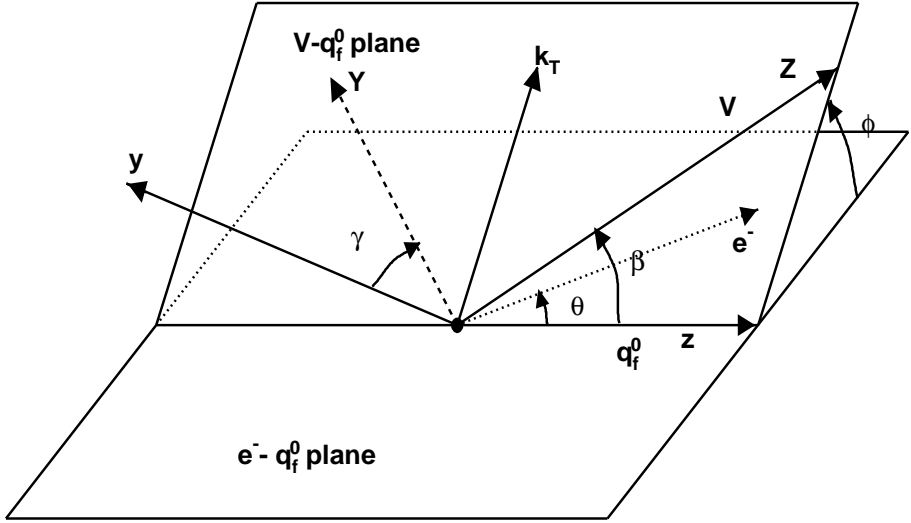,width=15cm}
\caption{Illustrating graph showing the relation between the helicity 
frame $oxyz$ of the initial quark $q_f^0$ and the helicity beam
frame $OXYZ$ of the produced vector meson $V$.
There are three planes involved here: the horizontal plane
determined by the incoming $e^+e^-$ and outgoing $q_f^0{\bar{q}}_f^0$,
the $x-z$ plane, i.e., $V-q_f^0$ plane determined by 
the moving direction of $q_f^0$ and that of V and 
${\vec e}_y={ {\vec{e}_z}\times {\vec{e}}_{k_\perp}}/
|{\vec{e}_z}\times {\vec{e}}_{k_\perp}|$,
the $X-Z$ plane determined by the moving direction 
of $q_f^0$ and that of the $e^-$ beam and
${\vec e}_Y={ {\vec{e}_Z}\times {\vec{e}}_{p_{e^-}}}/
|{\vec{e}_Z}\times {\vec{e}}_{p_{e^-}}|$. }
\end{figure}

\begin{figure}
\psfig{file=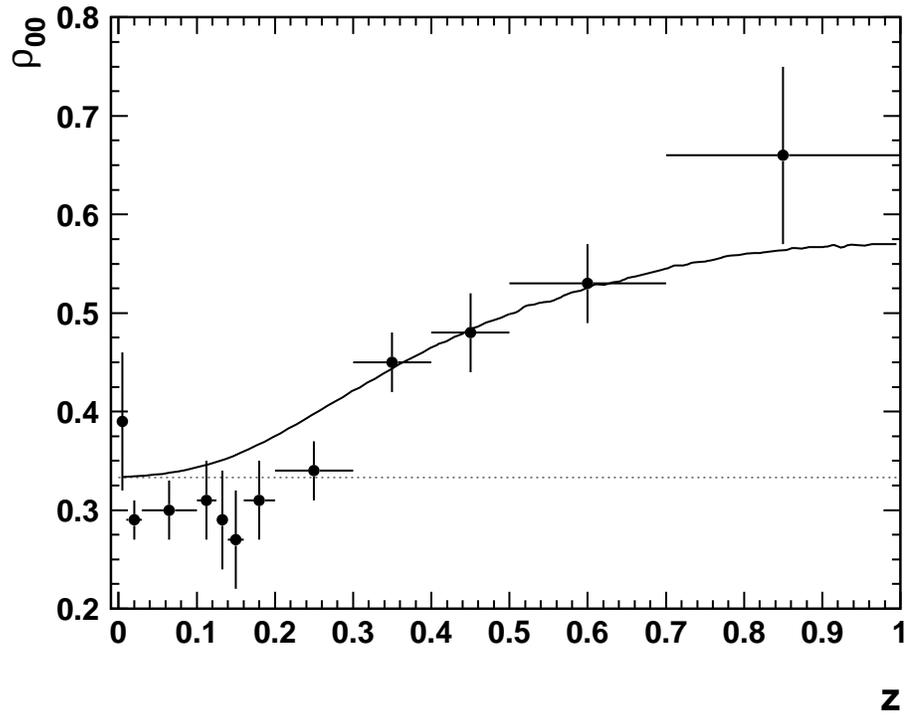,width=15cm}
\caption{Spin density matrix element $\rho_{00}$ of $K^{*0}$
produced in $Z^0$ decay obtained from Eq. (24) by taking $P_z=0.48$ 
as a function of the momentum fraction $z$. 
The dotted straight line corresponds to the unpolarized case (1/3);
the data are from Ref. [16].}
\end{figure}

\newpage

Figure caption:

\noindent
Fig. 1: Illustrating graph showing the relation between the helicity 
frame $oxyz$ of the initial quark $q_f^0$ and the helicity beam
frame $OXYZ$ of the produced vector meson $V$.
There are three planes involved here: the horizontal plane
determined by the incoming $e^+e^-$ and outgoing $q_f^0{\bar{q}}_f^0$,
the $x-z$ plane, i.e., $V-q_f^0$ plane determined by 
the moving direction of $q_f^0$ and that of V and 
${\vec e}_y={ {\vec{e}_z}\times {\vec{e}}_{k_\perp}}/
|{\vec{e}_z}\times {\vec{e}}_{k_\perp}|$,
the $X-Z$ plane determined by the moving direction 
of $q_f^0$ and that of the $e^-$ beam and
${\vec e}_Y={ {\vec{e}_Z}\times {\vec{e}}_{p_{e^-}}}/
|{\vec{e}_Z}\times {\vec{e}}_{p_{e^-}}|$. 

\noindent
Fig. 2: Spin density matrix element $\rho_{00}$ of $K^{*0}$
produced in $Z^0$ decay obtained from Eq. (24) by taking $P_z=0.48$ 
as a function of the momentum fraction $z$. 
The dotted straight line corresponds to the unpolarized case (1/3);
the data are from Ref. [16].

\vskip 1cm
Table caption:

\noindent
Tab. 1: Spin density matrix element $\rho_{00}$ 
for different vector mesons obtained from Eq.(25) 
using $P_z=-0.51P_f$.
The data are taken from Refs.[\ref{f4}-\ref{f3}].

\end{document}